\documentclass[useAMS,usenatbib]{mnras}
\usepackage{newtxtext,newtxmath}
\usepackage[T1]{fontenc}
\usepackage{ae,aecompl}
\usepackage{graphicx}	
\usepackage{amsmath}
\usepackage{amssymb}
\usepackage{subfigure}
\usepackage{color}

\usepackage{newtxtext,newtxmath}
\usepackage{amssymb, braket}
\usepackage{hyperref}
\usepackage{geometry}
\usepackage{enumitem}

\title[Cosmic Rays and the Chiral Puzzle of Life]{Cosmic Rays and the Chiral Puzzle of Life}
\author[]{No\'emie Globus$^{1,2}$ and Roger D. Blandford$^{3}$\\
$^{1}$Center for Cosmology \& Particle Physics, New-York University, New-York, NY10003, USA\\ $^{2}$Center for Computational Astrophysics, Flatiron Institute, Simons Foundation, New-York, NY10003, USA\\
$^{3}$Kavli Institute for Particle Astrophysics \& Cosmology (KIPAC), Stanford University, Stanford, CA 94305, USA}
\date{Released \today}

\pagerange{\pageref{firstpage}--\pageref{lastpage}} \pubyear{2002}

\def\LaTeX{L\kern-.36em\raise.3ex\hbox{a}\kern-.15em
    T\kern-.1667em\lower.7ex\hbox{E}\kern-.125emX}

\begin{document}
\label{firstpage}
\maketitle
\begin{abstract}
Living organisms exhibit consistent homochirality. It is argued that the specific, binary choice made is not an accident but is a consequence of parity violation in the weak interaction expressed by cosmic irradiation. The secondary muons and pairs are spin- and magnetic moment- polarized and may introduce a small, net chiral preference when they interact electromagnetically or quantum mechanically with molecules that have made the transition to self-replication. Although this preference is likely to be very small, it may suffice to give a chirally-dominant outcome after billions of replications, especially if combined with chirally-unbiased conflict between the two choices. Examples of mechanisms that can manifest the three essential steps of polarization, preference and domination are presented and some variations and possible implications are discussed.
\end{abstract}

\maketitle

\section{Introduction}\label{sec:intro}
Living organisms comprise a system of molecules organized with specific handedness. The ribonucleic and deoxyribonucleic acids (RNA and DNA) are responsible for the replication and storage of genetic information. Both are made up of linear sequences of nucleotides that encode the linear sequence of amino acids in a specific protein. DNA is a double helix, while RNA is a single helix. The helix is an  extremely common structural feature in biomolecules; it is a general response to the pile up of single monomer units into a polymer. (By monomers we refer to the helix components, {\textit i.e.}, the  nucleotides in DNA or the amino acids in proteins). The helix appear to be left- or right-handed, depending on the handedness of the monomers. Biology on Earth has made one of the two choices of opposite handedness,  
{\textit i.e.}, right-handed sugars and left-handed amino-acids that lead to  predominantly right-handed helical configurations in biopolymers \citep{Biswas16}.
 In what follows, we eschew the words ``left'' and ``right'' in describing structure and polarization because they have been used differently, inconsistently and, consequently, confusingly over the many subfields that contribute to our discussion; instead, we denote the (finally) chosen set of molecules as ``live'' and the alternative set as ``evil''. The live choices appear to be sustained, once they are established, because the presence of single,  evil monomers in  live helices can de-stabilize them. 

This apparent choice of handedness is often called "chirality" \citep{Kelvin1894}. This word has taken on several subtly different meanings which should be distinguished. To Kelvin, chirality was a geometric property of a set of points in three dimensional space that could not be translated and rotated to coincide with its inversion about a point. Biological chirality is different and can encompass a larger set of phenomena than what concerns us here, for the example the helical growth of some creeping plans that can be induced by the rotation of the earth relative to the direction of the sun in one hemisphere. However, most biological chirality is believed to be derivative of the underlying chemical chirality of the majority of biotic molecules. Here, Kelvin's points are replaced by atoms connected by single sigma bonds that can allow relative rotation of large groups. Kelvin's definition can also apply to a crystal which is a geometric arrangements of molecules. Most importantly, it applies to the arrangement of bases in a DNA molecule which is neither periodic nor random and contains the genetic information needed to sustain life \citep[{\textit e.g.}][]{Schrodinger,Shannon48, WatsonCrick, Shinitzky07}.

These simple, helical and consistent systems presumably originated around the same time as the transition from chemical reactions between pre-biotic molecules to self-replication and evolution of the earliest and simplest biological molecules --- in an environment, that we call the ``fount'' of life. This fount might be found on a young Earth, another planet, a satellite, an asteroid or a comet. Since the pioneering works of \citet{Miller}, and \citet{Oro60,Oro61},   many experiments have  demonstrated how to synthesize amino acids and DNA bases, by irradiation of mixtures simulating  our primitive atmosphere or the interstellar medium \citep[see][for a review of the different experiments related to the synthesis the building blocks of life]{Kitadai}. Ice seems to be more favorable to RNA replication than a super-cooled solution of the same temperature \citep{Attwater}. The presence of phyllosillicates is considered to be an important factor in prebiotic chemistry \citep[][]{Erastova}. \citet{Ruff04} showed spectral evidence for zeolite in the Martian dust; and recently, \citet{Ruf17} reported the finding of new metallo-organic compounds in meteorites. Zeolites are generally considered to be a family of materials with highly symmetrical structures, but many of them are chiral and capable of enantioselective recognition \citep[][]{Dryzun}. Such a selection might have helped to form the first chiral domains of simple monomers needed to assemble the first helical biomolecules.  Direct evidence of complex prebiotic chemistry is provided by the 4.5-billion-year-old organic materials hosted in the most primitive and least altered meteorites \citep{Gilmour2003}. Some of these organic molecules come in two forms that are mirror images of each other. Those molecules are called "enantiomers", from the Greek $\epsilon\chi\theta\rho$\'o$\varsigma$, "enemy" \citep{Meierhenrich}. It was found that in some meteorites, one of the two forms is present in greater quantity than the other \citep{Pizzarello2006}. These enantiomeric excesses are an important clue to the role that meteorites could have played in the origin of life on Earth. 

Chirality is pretty much inevitable in organic chemistry because of the peculiar atomic properties of carbon atoms with four second shell electrons that can form three of four non-coplanar and different bonds. Typically a chiral carbon is a carbon with four distinct atoms or groups attached to it \citep{vantHoff74,LeBel74} (the carbon is $sp^3$ hybridized). No other atom can enter into four robust and versatile bonds and, for this reason, atomic life forms are generally argued to be stable. However, this allows two sets of enantiomers to develop along separate synchronized paths making similar evolutionary choices in response to changes in common environments.This clearly did not happen and a single choice was made, with an entropic price that is surely paid by the greater facility of storing information and the higher reliability of the replication \citep[\textit{c.f.}][]{Schrodinger}. A precise equilibrium between the two chiral choices seems quite unlikely, given the high replication rate, if we accept the inevitability of homochirality, wherever life is found, at least so far. The essential point is that we can imagine a biology that makes the evil choice consistently. In this alternative biology, everything would function in the same way except for very small effects that are the main topic of this article. Either there was a unique spatial and temporal source for life where a chirality choice was made, probably randomly, and all expressions of it have subsequently adapted and dispersed retaining the initial choice, or the live choice was made for a reason and preserved deterministically everywhere. This allows life to originate independently at many different sites and epochs with the same chirality. As there could be billions of generation of the earliest and simplest life forms, a small bias could easily ensure and sustain homochirality. 

This brings up physical chirality. \citet{Pasteur48} discovered biological chirality when he found that tartaric acid derived from a living source rotated the plane of linearly polarized light. This means that circularly polarized eigenmodes propagate with slightly different speeds. By contrast, tartaric acid derived from chemical synthesis did not behave in this fashion. However, Pasteur was able to separate crystals of synthesized tartaric acid into two groups on the basis of their geometrical shape and crystals from one group behaved like the biological material and those from the other group rotated the plane of polarization in the opposite direction. In this way, Pasteur discovered biological homochirality and recognized it as a consequence of some asymmetry in the laws of nature: {\it "If the foundations of life are dissymmetric, then because of dissymmetric cosmic forces operating at their origin; this, I think, is one of the links between the life on this earth and the cosmos, that is the totality of forces in the universe"} \citep[][for the translation]{Quack89}. 

Had Pasteur been alive a century later, the discovery of parity (P) violation in the weak interaction \citep{LeeYang1956,Wu1957} would have strengthened his view. A parity operation, P, can be thought of as a reflection in a mirror plane or, equivalently, inversion through a point. Parity violation is a signature of the weak interaction, in contrast to strong and electromagnetic interactions. Fermions must have ``left-chirality'' to interact with a $W$ boson through the weak interaction and their antiparticles must have ``right-chirality''. In the limit that a lepton is massless, the term "chirality" coincides with helicity (the projection of the spin vector on to their momentum vector). In the original, standard model of particle physics, the neutrino is massless. Although the discovery of small neutrino masses implies that helicity is strictly frame-dependent, this is unimportant for biological and chemical purposes and neutrinos are effectively chiral particles. By contrast, gravitational, electromagnetic and strong interactions are symmetric under parity change and, so, a physics-based, causal explanation of biological chirality is almost sure to involve the weak interaction.  

\begin{figure*}\label{fig:demeter}
\centering
\includegraphics[scale=0.27]{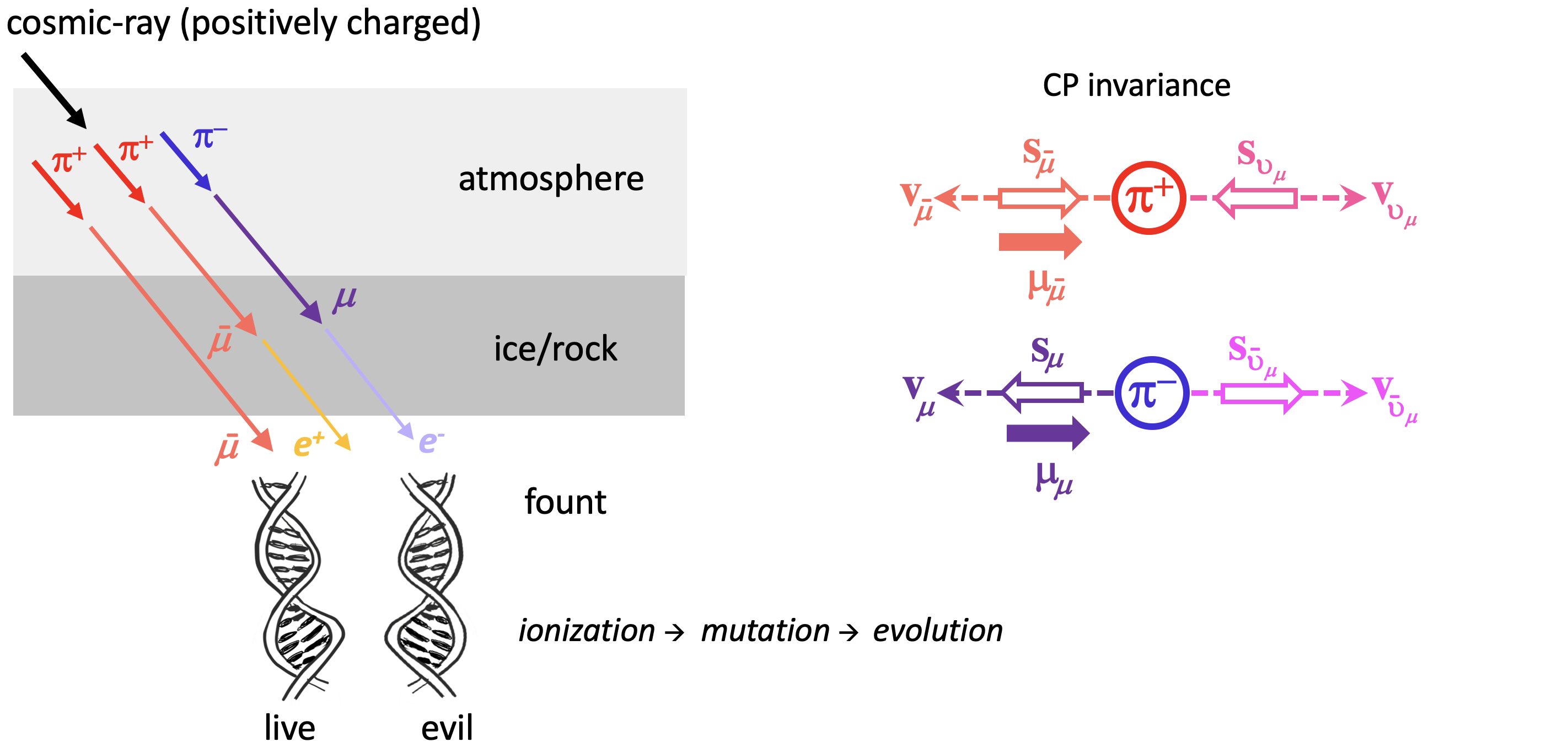}
\caption{Left: spin-polarized muons and electrons (respectively antimuons and positrons) from charged pion decay in extensive air showers. The cosmic-ray lodacity ${\cal L}=<{\hat{\boldsymbol{\muup}}\cdot\hat{\textbf v}}>$, always negative, is a consistent  chiral environmental factor. Other environmental factors, like an external magnetic field or a fount that might oriente the molecules, would involve ${\cal M}_{\rm e}$ and lead to a larger chiral bias. Right: CP invariance in the pion decay that leads to an universal   sign of the cosmic-ray  lodacity ${\cal L}<0$. }
\label{demeter}
\end{figure*}

Enantiomers are truly chiral, in Kelvin's sense, and their basic Hamiltonian does not commute with the parity operator. If we add weak neutral currents to the Hamiltonian, there is a small energy difference (the Parity Violating Energy Difference, hereafter PVED), equal in magnitude but opposite in sign between enantiomers, making one molecule more stable than the other \citep{Yamagata66}. Calculation of PVED in aqueous solution indicates a preferred stability of left-handed amino acid and right-handed sugars, as observed in living organisms. However, this difference is extremely small, PVED/kT $\sim 10^{-17}$. It was concluded that a collective effect, perhaps involving spontaneous symmetry breaking, would be needed to explain homochirality. \citet{Salam91,Salam92} proposed that a phase transition into the more stable enantiomeric form occurs at a certain critical temperature. This has not been validated  by laboratory experiment.   

In a beautiful paper, Pierre Curie  addressed the question of  chirality transfer from light to molecules, specifically involving circular polarisation \citep{Curie94}.  As noted above, the first indication of handedness was the observation that the refractive indices for two circularly-polarized normal modes of propagation of light could be slightly different, resulting in a rotation of the plane of polarisation of incident linearly polarized light \citep{Pasteur48}. The sense of the rotation reflects the underlying chirality of the molecules, though the relationship is not simple and depends upon the wavelength of the light (Optical Rotatory Dispersion). This rotation can be accompanied by a difference in the absorption (Circular Dichroism), consistent with the \citet{Kramers,Kronig} relations. On this basis, it has been suggested that a specific source of circularly polarized light might favour one set of enantiomers over the other \citep{Bailey98}.  

Laboratory experiments have demonstrated that it is possible to induce an enantiomeric excess of amino acids  by irradiation of interstellar ice analogs with UV-circularly polarized light (UV CPL) \citep{deMarcellus}. However, this raises two problems. Firstly, Circular Dichroism is also wavelength, pH and molecule specific \citep{Hendecourt}. It is hard to see how one sense of circular polarization can enforce a consistent chiral choice, given the large range of conditions under which the molecules are found. Secondly, it is often supposed that astronomical sources supply the polarization. However, optical polarimetry within the Galaxy reveals no consistent sense of circular polarisation and the observed degrees of polarization are quite small \citep{2001OLEB...31..167B}. This can still only produce a local chirality as most of the light sources are achiral. CPL in the near infrared has been observed up to 20\% in star forming regions \citep{Kwon}. However the level of CPL is related to the amount of extinction and scattering, and will be less in the UV. If we seek a  universal, chiral light source, that consistently emits one polarization over another, then we are again drawn to the weak interaction in order to account for a universal asymmetry. One option is to invoke spin-polarized particles, which can radiate one sense of circular polarization through {\v C}erenkov radiation or bremsstrahlung and can preferentially photolyze chiral molecules of one handedness \citep{VesterUlbricht59,Lahoti77}. However, the resulting chiral differences are again very small. This suggests considering, instead, the direct interaction of the particles themselves with biological molecules.

Cosmic rays are a source of polarized particles. Probably the most relevant cosmic-ray sources are the solar corona and hydromagnetic disturbances in the solar wind. These produce, mostly, mildly relativistic protons. 
The young sun and its wind, to which early life was exposed, are likely to have been much more active then because the sun presumably rotated faster. 
Galactic cosmic rays mostly produced by supernova remnants could also be important but they are accelerated with a much harder spectrum, which is further hardened by modulation by the stellar wind but are likely to be sub-dominant.

With reference to conditions on Earth, today, we are mostly interested in charged protons, with energies just above the threshold for pion production, colliding with nitrogen and oxygen nuclei in the upper atmosphere (Fig.~\ref{fig:demeter}). The resulting pions, mostly $\pi^+$, created through the strong interaction, undergo weak decay within a few meters into (positive) anti-muons with half lives $\sim2\,\mu{\rm s}$, which decay, in turn, also weakly, into (positive) positrons. A minority of $\pi^-$ create muons and electrons. Most cosmic rays lose energy though ionizing electrons from air molecules and they and the  daughter particles are stopped in the upper atmosphere. However, a significant number of muons do make it down to ground level and some even penetrate deep underground, before they decay. In this way, the atmosphere can serve as an effective $\mu$ filter. (We emphasize that the conditions in the young Earth and in other locales are likely to be qualitatively different and may have to be specialized to induce homochirality.)  

As the weak interaction is involved and pions are spinless, the anti-muon and positron spins, \textbf s, are preferentially anti-aligned with their velocity, $\textbf v$ in order to balance the antiparallel spins of the accompanying antineutrinos (Fig.~\ref{fig:demeter}). By contrast, the muon and electrons spins from negative pion decay would be preferentially aligned with their velocity. The associated magnetic dipole moments are given by $\boldsymbol{\muup}=\gammaup \textbf s$ where $\gammaup=g(e/2m)$ is the gyromagnetic ratio. As they have positive (negative) charge, then $g\approx2(-2)$ for the antimuons and positrons (muons and electrons). As a consequence, all of these particles have magnetic moments preferentially anti-aligned with their velocity. If we introduce the operation of charge conjugation, C, this is is an expression of CP conservation. However the magnitude of the magnetic moment is $m_\mu/m_e\sim200$ times smaller for $\mu^\pm$ than for $e^\pm$.  

Now the $\mu^\pm,e^\pm$ will also undergo ionization energy loss and this will diminish the alignment. It is normal to focus on the helicity polarization which describes the average component of spin along the velocity. However, in this paper, we are mostly concerned with the magnetic moments of these particles and so we introduce a quantity which we call ``lodacity'' (after lodestone) defined by 
\begin{equation}\label{eq:lodacity}
{\cal L}(T)=<\hat{\boldsymbol{\muup}}\cdot\hat{\textbf v}>,
\end{equation}
where we average over all cosmic rays of a given type and kinetic energy $T$. The magnetic moment $\boldsymbol{\muup}$ includes the gyromagnetic ratio and therefore the lodacity has the same sign for the charge conjugate reaction, even if the helicity is opposite (see Fig.~\ref{fig:demeter}). ${\cal L}\gtrsim-1$ for freshly created $\mu$, and ${\cal L}\gtrsim-0.3$ for new $e$. However, when we consider the particles that irradiate biological molecules, ${\cal L}$ may be much smaller (though still negative) because the magnetic moment polarization of the original particles will have been degraded and there may also be a background of additional, unpolarized, secondary particles. 

Cosmic rays have a vital role. At modest intensity, which interests us, they promote mutation and natural exploration of biochemical and evolutionary pathways; when the intensity is high, they will be destructive and will create sterile environments. Now, when a cosmic ray ionizes a biological molecule, there is a small probability that it will liberate an electron, through classical Coulomb scattering. Other, bound transitions are possible and can be very important \citep{Rosenfeld28}. However, it is simple ionization that we will choose as a proxy for cosmic-ray induced mutation and a demonstration of chiral preference.

At this point, we should discuss how to describe a chiral molecule or, more generally, a chiral unit - a small portion of a much larger molecule that can have significant interaction with an individual cosmic ray. We suppose that it is characterized by another pseudoscalar quantity, which we call the molecular chirality $\cal M$. This has opposite signs for live and evil molecules. Now, the chiral unit comprises a small number of nuclei surrounded by electrons with mean electron density reflecting the disposition of inner shells and chemical bonds. If we try to capture the non-spherical charge distribution in terms of a spherical harmonic expansion, either semi-classically or quantum mechanically, then we can introduce an electric dipole moment vector, \textbf d, and a (trace-free) electric quadrupole tensor, \textsf Q. In addition, the electrons in the chiral unit possess orbital and spin angular momentum and this can create a magnetic dipole moment \textbf m. Treated as operators, $\textbf d$ anticommutes with P, while $\textbf m$ and $\textsf Q$ commute. The simplest type of molecular chirality we can describe, and which we will emphasize, is ``electromagnetic chirality'' 
\begin{equation}\label{eq:molchir}
{\cal M}_{\rm em}=\hat{\textbf d}\cdot\hat{\textbf m}.
\end{equation}
If this is non-zero and can couple to $\cal L$, then there is a chiral preference. There is also the possibility of electric molecular chirality, involving just \textbf d and \textsf Q. The quadrupole tensor has three orthogonal eigenvectors, each of either sign. Alone, this cannot be  chiral. However, if there is also an electric dipole moment lying within one of the octants then it chooses the three eigenvectors with which it makes an acute angle and the triple product of these three eigenvectors, taken in the order of their size, defines a pseudoscalar, the  ``electric chirality'' 
${\cal M}_{\rm e}$, which can also, in principle, couple to the cosmic rays\footnote{``Magnetic'' chirality, only involving magnetic field, can arise in ferromagnetic material \citep[{\textit e.g.}][]{Grigoriev} but need not concern us here.}.  

Cosmic-rays provides a natural  connection between the weak interaction and living systems. Because of the lodacity which is always of the same sign, the cosmic radiation assure small but permanent bias (reflected by the sign of the product $\cal{M}\cdot\cal{L}$ which is always different for live and evil molecules) that can lead to a {\it global} symmetry-breaking as anticipated by Pasteur.

In this paper we propose that pre-biotic chemistry produces both live and evil versions of the molecular ingredients of life. At some stage in the earliest development and evolution of living, specifically copying and reproducing, molecules a small difference in the ionization/mutation rate, attributable to the spin polarization of the cosmic rays, gives a chiral preference to the live molecules over their evil counterparts. To be more specific, in the simplest example the ionization rate, a scalar, contains contributions that contain the product $\cal M\cdot\cal L$. Given the large number of generations of the simplest living organisms, a small preference suffices to lead to chiral dominance of live molecules. (Actually, as we shall discuss, the dominance can assert itself quite quickly if there is also a competition between live and evil molecules.) We suspect that this mechanism can only be relevant for the simplest organisms; more complex forms of life will involve a larger range of more powerful evolutionary selection effects.  

The plan of the paper is as follows: in Section~\ref{sec:cosmicrays}, we summarize the properties of the air showers and review the different spin-polarized components. We show the chiral propagation from cosmic-rays to biomolecules with a simple model in Section~\ref{sec-chiralpropa}. Then we discuss our evolutionary scenario that leads to homochirality in the Section~\ref{sec-evolampli}. We conclude with a short discussion of variations and further implications in Section~\ref{sec-discussion}.

\section{Cosmic rays}\label{sec:cosmicrays}
\subsection{Background levels and mutagenesis}
Let us now examine the general properties of cosmic ray showers, and their biological consequences. As recalled in section \ref{sec:intro}, cosmic rays are continuously hitting the Earth's atmosphere inducing extensive air showers (EAS) by successive interactions with the air molecules. EAS are not only produced in the planet's atmospheres but also in the mantles of comets and other icy bodies that are exposed to solar and Galactic cosmic rays. Cosmic rays are a necessary factor in life's evolution. A background level of radiation seems to  stimulate living systems. By contrast, radiation deprivation  inhibits bacterial growth \citep{Planel87,Kawanishi2012,Castillo15}. It has also been shown that during episodes of high cosmic-ray flux and cold climate there is an enhancement of biological productivity \citep{Svensmark}.  

Some fraction of environmental mutagenesis during  the evolution of life is attributable to ionization by cosmic rays. Radiation increases the frequency of gene mutations; this is known since the pioneering work of \citet{Muller} that showed that the mutation rate is proportional to the radiation dose. DNA is the most critical biological target because of the genetic information it contains. Damage to DNA may be expressed in the form of mutations, the frequency of which appears to increase as a linear function of the particle flux, $M=\sigma_M F$, where $\sigma_M$ is the mutation induction cross section (in $\mu m^2$), $F$ is the particle flux (in $\mu m^{-2}s^{-1}$). An empirical approximation for the mutation cross section is given by \cite{Kiefer}, \begin{equation}\sigma_M=\sigma_0[1-\exp(-3.68\,10^{-3}L/\sigma_0-7.\,10^{-5}L^2)]\,,
\label{mutation_cross_section}
\end{equation}
where $L$ is the Linear Energy Transfer and $\sigma_0\sim10^{-5}\mu m^2$.  $L$ is used to characterize biological damage; it is the average amount of energy deposited per unit length of the substance ($L = dE/dx$ in keV/$\mu m$). It has been shown that the biological damage is roughly proportional  to the  muon flux, and that the fluence-to-dose ratio remains fairly constant with energy \citep{Chen2006, Pelliccioni2000}. 

On Earth, the  overall globally averaged annual radiation dose from natural sources is 2.4 mSv/yr, the cosmic ray component is 0.39 mSv/yr and the muon component is 0.33 mSv/yr \citep{Atri2011}. This is because the muon component dominates the flux of particles on the ground at energies above 100 MeV, contributing to 85\% to the radiation dose from cosmic rays. Muons arrive at sea level with an average flux of about 1 muon per square centimeter per minute. Muons typically lose almost 2~GeV in passing through the atmosphere ($\sim$2~MeV g$^{-1}$cm$^{-2}$). Muons are the only biologically significant cosmic radiation with energy sufficient to penetrate considerable depths, and they are, on average, spin-polarized. The mean energy of muons at the ground under contemporary conditions is $\sim4$~GeV which is enough to penetrate a few meters of rock and several hundred meters of ice.

\subsection{Air shower asymmetries}\label{ASA}
\subsubsection{Charge ratio}\label{charge_ratio}
Primary cosmic rays comprise mostly positive nucleons. This excess is transmitted via nuclear interactions to pions and then, on to muons.  The muon charge ratio is ${\cal R}_\mu\sim1.25$ below 1 TeV and increases to above $\sim1.4$ at higher energies \citep{Gaisser}.  

Due to parity violation in the weak interaction, $\mu^\pm$ produced from decaying pions and kaons are on average spin-polarized. (The dominant contribution is from pion decay.) Their daughter electrons and positrons are also, on average, spin-polarized. The spin-polarized cosmic-rays can also produce UV CPL when propagating in the medium through emitting {\v C}erenkov radiation and bremsstrahlung.  

\subsubsection{Spin-polarized secondary particles}\label{spin-polarizedCR}
The spinless charged pion with a lifetime of 26 ns decays at rest into a left-handed muon neutrino and a muon:  $\Pi^-\rightarrow \mu\bar{\nu}_\mu$ (and $\Pi^+\rightarrow \mu^+\nu_\mu$ respectively). 
The pion has a mass of $m_\pi$ = 140 MeV/c$^2$, the muon has a mass $m_\mu$ = 106 MeV/c$^2$ and the neutrino is considered massless.
We define $r_\pi=(m_\mu/m_\pi)^2$.
In the pion rest frame (denited by $*$), the momentum of the muon is 
\begin{equation}|\textbf p^*_{\mu}|=|\textbf p^*_{\nu}|=\frac{m_\pi c}{2}(1-r_\pi)\sim 29.8\, {\rm MeV/c} \end{equation}
and $E^*_\mu=\sqrt{{p^*_\mu}^2 c^2+m_\mu^2c^4}\sim109.8$ MeV.
The velocities of the muon and neutrino in the rest frame of the pion are $\beta^*_\mu=p^*_\mu/E^*_\mu\sim0.271$ and $\beta^*_{\nu_\mu}=1$.
Let the pion move in the laboratory with velocity $\textbf v_\pi/c={\beta}_\pi \textbf{e}_z$. 
Defining $\theta^*$, the angle of emission of the muon in the pion rest frame, we have the following relations for the muon momentum, energy, helicity ($h = \hat{\textbf{s}}\cdot\textbf{p}/|\textbf{p}|$) and angle of emission in the lab rest frame \citep{Lipari93}:
\begin{eqnarray} 
p_\mu=\gamma_\pi{p^*_\mu}\cos\theta^*+\beta_\pi\gamma_\pi{E^*_\mu}\,,\\
E_\mu=\gamma_\pi{E^*_\mu}+\beta_\pi\gamma_\pi{p^*_\mu}\cos\theta^*\,,\\
h(\beta_\pi,\theta^*)=\frac{1}{\beta_\mu}\left[ \frac{1-r_\pi+(1+r_\pi)\cos\theta^*\beta_\pi}{1+r_\pi+(1-r_\pi)\cos\theta^*\beta_\pi}\right]\,,\\
\tan\theta=\frac{\beta^*_\mu\sin\theta^*}{\gamma_\pi(\beta_\pi+\beta^*_\mu\cos\theta^*)}\,.
\end{eqnarray}
In the limit $\beta_\pi = 0$, the velocity of the muon is: $\beta_\mu = (1 - r_\pi)/ (1+r_\pi)\equiv\beta^*_\mu\sim0.27$  and we have $h$ = + 1 independently from the angle of emission of the muon. The polarization of the positive muon flux at sea level varies between $\sim30$\% and $\sim60$\%, depending on the energy, and is higher than the polarization of the negative muon flux \citep{Lipari93}. The lifetime of negative muons in matter is different because the negative muons interact with the nuclei of atoms, which will increase the charge ratio at greater depth. 
In the same fashion, the electrons (and respectively positrons) from muon decay (antimuon decay) are mostly left-handed (right-handed) with the direction of the spin-aligned (opposite) to their momentum: $\mu^-\rightarrow e^-\nu_\mu\bar{\nu}_e$ ($\mu^+\rightarrow e^+\nu_e\bar{\nu}_\mu$).
The decay probability of a positron is $W(\theta)=(1+a\cos\theta)/(4\pi\tau_\mu)$ where $\theta$ is the angle between the spin direction and the positron trajectory, $\tau_\mu \sim 2.197 \,\mu$s is the mean lifetime, and the asymmetry term $a$ is a direct consequence that the muon decay is governed by the weak interaction, and depends on the positron energy, so the positron  angular distribution is ${\rm d}\Gamma/{\rm d}\cos\theta= W(\theta)$. The maximum and mean positron energies resulting from the three body decay are given by: $E_{{e^+}_{\rm max}}=(m_\mu^2+m_e^2)c^2/(2m_\mu^2)= 52.82$ MeV and $\bar{E}_{e^+}=36.9$ MeV. For a positron emitted with  energy of the order of $E_{{e^+}_{\rm max}}$, we have the maximum asymmetry $a=1$.   When averaged over all positron energies, $a=1/3$.  
\subsubsection{Circularly polarized radiation}
{\v C}erenkov radiation has a degree of polarization which is dependent on the orientation of the spin of the initial particle. 
This is a purely relativistic quantum effect \citep{Sokolov40}.
In the following we consider the difference between the number of left-handed photons and right-handed photons emitted from an electron of helicity $-1/2$ and velocity $\beta\sim0.8$ propagating in ice. 
Defining the ratio of the photon to the electron energies, $\xi= {\hbar\omega}/(2E_e)$, the velocity and Lorentz factor of the electron $\beta=v/c$, $\gamma=(1-\beta^2)^{-1/2}$,
the {\v C}erenkov angle $\cos\theta_c=[1+\xi(n^2-1)]/(n\beta)$,
 and the function ${\cal F}=\cos\chi(\cos\theta_c-n\beta)+\gamma^{-1}\sin\chi\cos\phi\sin\theta_c$
where the angles are $\alpha=\pi/4$, $\chi=0$, $\phi=0$ for circular polarization,
the number of right-handed photons $N_{1,+}$ (respectively left-handed $N_{1,-}$) is \citep{Lahoti77}
\begin{align}
N_{1,\pm}\propto 0.5(\beta\sin\theta_c)^2+\xi^2(n^2-1)\pm0.5(\beta\sin\theta_c)^2\cos(2\alpha)\nonumber\\\mp\sin(2\alpha)\xi {\cal F}\,.
\end{align}
As an example, the ratio $(N_{1,+}-N_{1,-})/(N_{1,+}+N_{1,-})$  emitted by an electron  of energy $\sim0.8$ MeV ($\beta=0.77$), propagating in ice, is $\sim 1.3\,10^{-5}$ at a wavelength of 206 nm. For muons at the same velocity ($\sim166$ MeV), the ratio is $1.8\,10^{-8}$ at the same wavelength. 

Longitudinally polarized $\beta$-radiation gives rise to circularly polarized bremsstrahlung. 
Using the Born approximation, \citet{McVoy57} derived the following formula for circular polarization in the limit where $E_e\sim h\nu$ and the emission angle of the photon $\theta=0^\circ$:
\begin{equation}
\frac{P_{\gamma}}{P_e}=\left(1+\frac{(1-\beta)(E_e+2mc^2)}{(2-\beta)E_e}\right)^{-1}\,.    
\end{equation}
Here $P_e$ is the polarization of the electron. 
The polarization transfer drops rapidly at electron energies $E_e$ below $\sim$1 MeV.

\subsubsection{Evolution of Spin Polarization}
\label{sssec:ESP}
As we discussed in the previous section, cosmic rays are preferentially positively-charged and create 
$\mu^+$ and $e^+$ with mean magnetic moments $\boldsymbol{\muup}$ anti-parallel to the velocity. This asymmetry can be degraded by two effects. The first is precession about an external magnetic field, $\textbf B_{\rm ext}$ \footnote{Precession within the molecule is ignorable.}; the second is deflection of the particle momentum in a Coulomb interaction that also leads to energy loss while leaving the direction of the magnetic moment unchanged. We consider these, in turn, for antimuons/muons and for positrons/electrons.

Muons are unmagnetized (their Larmor radius exceeds the length of their trajectory) so long as $B_{\rm ext}\lesssim1\,{\rm mT}$. 
By contrast, positrons and electrons are magnetized so long as they traverse a length $L\gtrsim(p/(m_{\rm e}c))(B_{\rm ext}/1\,{\rm mT})^{-1}\,{\rm m}$, where \textbf p is the electron Lorentz factor, a condition that is quite likely to be satisfied.

The equation of motion for a magnetized positron is 
\begin{equation}
\frac{d\textbf p}{dt}=\frac e{\gamma m_{\rm e}}\textbf{p}\times\textbf B_{\rm ext},
\end{equation}
where $\gamma=(1-p^2/m_{\rm e}^2c^2)^{-1/2}$. The magnetic moment will also precess about the magnetic field according to
\begin{equation}
\frac{d\boldsymbol{\muup}}{dt}=\frac e{\gamma m_{\rm e}}\boldsymbol{\muup}\times\left(\textbf B_{\rm ext}+\frac{(\gamma-1)(\textbf B_{\rm ext}\cdot\textbf v)\textbf v}{v^2} \right).
\end{equation}
In the non-relativistic limit, which concerns us most, these equations then imply that \textbf p and $\boldsymbol{\muup}$ precess about $\textbf B_{\rm ext}$ with a common angular velocity $-eB_{\rm ext}/m_e$. We expect the spin-polarized daughter positrons to outnumber spin-polarized electrons of similar momenta and to be created with a momentum distribution that is axisymmetric about a downward direction, $\hat{\textbf g}\equiv g\hat{\textbf e}_z$. Furthermore, for each \textbf p, the distribution of $\boldsymbol{\muup}$ will be axisymmetric about $\textbf B_{\rm ext}$. For a given magnetic field direction, this can lead to an average spin/magnetic moment polarization projected perpendicular to the velocity. However, in this case, it is only $\cal L$ that has the required pseudoscalar form and the perpendicular component leads to no bias after full averaging. The precession contributes modest degradation of the mean polarization. 

The influence of deflection can also be discussed. We consider all the spin-polarized cosmic rays from pion decay, starting with the same energy. We describe here the  evolution of the mean polarization (or equivalently lodacity) as they decelerate mainly by Coulomb scattering distant electrons.
For a single cosmic-ray, the spin direction in space should not change in a single encounter but the the velocity will undergo a deflection through a small angle, with the recoiling electron removing kinetic energy from the cosmic ray. We can regard this is a diffusion of the angle $\theta$ the spin makes with the velocity. As distant encounters dominate, we find that the diffusion coefficient satisfies 
\begin{equation}
\frac{d(\Delta\theta)^2}{d\ln\tau}=4D=1
\end{equation}
where $\tau=1+2m_ec^2/T$ \citep[\textit{e.g.}][]{ThorneBlandford17}.
The probability distribution per unit solid angle $P(\theta,\tau)$ satisfies the diffusion equation
\begin{equation}
\frac{\partial P}{\partial\ln\tau}=\frac1{\sin\theta}\frac\partial{\partial\theta}\sin\theta D\frac{\partial P}{\partial\theta}.
\end{equation}
Multiplying by $\cos\theta$ and integrating over solid angle, we obtain 
\begin{equation}
\frac{d\ln<\muup_z>}{d\ln\tau}=-\frac12,
\end{equation}
so that $<\muup_z>\propto\tau^{-1/2}$. 

Now consider the evolution of the mean spin polarization of all secondary cosmic-rays of same mass $m_e$. If the cosmic-rays are created relativistically with lodacity ${\cal L}_0$, then 
\begin{equation}\label{eq:LodT}
{\cal L}(T)\sim\,{\cal L}_0\left(\frac T{2m_ec^2}\right)^{1/2}.
\end{equation}

\section{Chiral Transfer from Spin-Polarized Radiation to Biomolecules}\label{sec-chiralpropa}
\subsection{Molecular Model}
\subsubsection{Chiral Carbon Atom}
In order to understand how spin-polarized cosmic ray showers might favor one chirality over the opposite choice, we make a very simple, semi-classical model that captures little of the actual biological, chemical and physical complexity. However, it is sufficient for our purpose as it incorporates the minimal and generic elements required to exhibit a chiral bias. In addition, it provides a useful guide for setting up a more realistic, general calculation and helps identify special conditions that might lead to larger effects.

We suppose that the principal chiral unit is the Carbon atom. We further suppose that Carbon atoms typically share eight electrons with four, non-coplanar, and different neighbors. We then localize four of these bonding electrons on a sphere of radius $R\sim100\,{\rm pm}$, roughly two thirds the length of a single Carbon bond, and ignore the neighbouring atoms including their electrons. The four bonding electrons then screen out the Carbon nuclear charge, well beyond the sphere. We suppose that these four electrons share a common, mean, independent ionization potential, $I\sim0.5\,I_H$, where $I_H\sim13.6\,{\rm eV}$ is the first Hydrogen ionization potential. We have explored a more realistic alternative model that includes adjacent base pairs drawn from a long helix but a single chiral unit contains the essential properties and is simpler to describe.

\begin{figure}\label{fig:quadchir}
\centering
\includegraphics[scale=0.19]{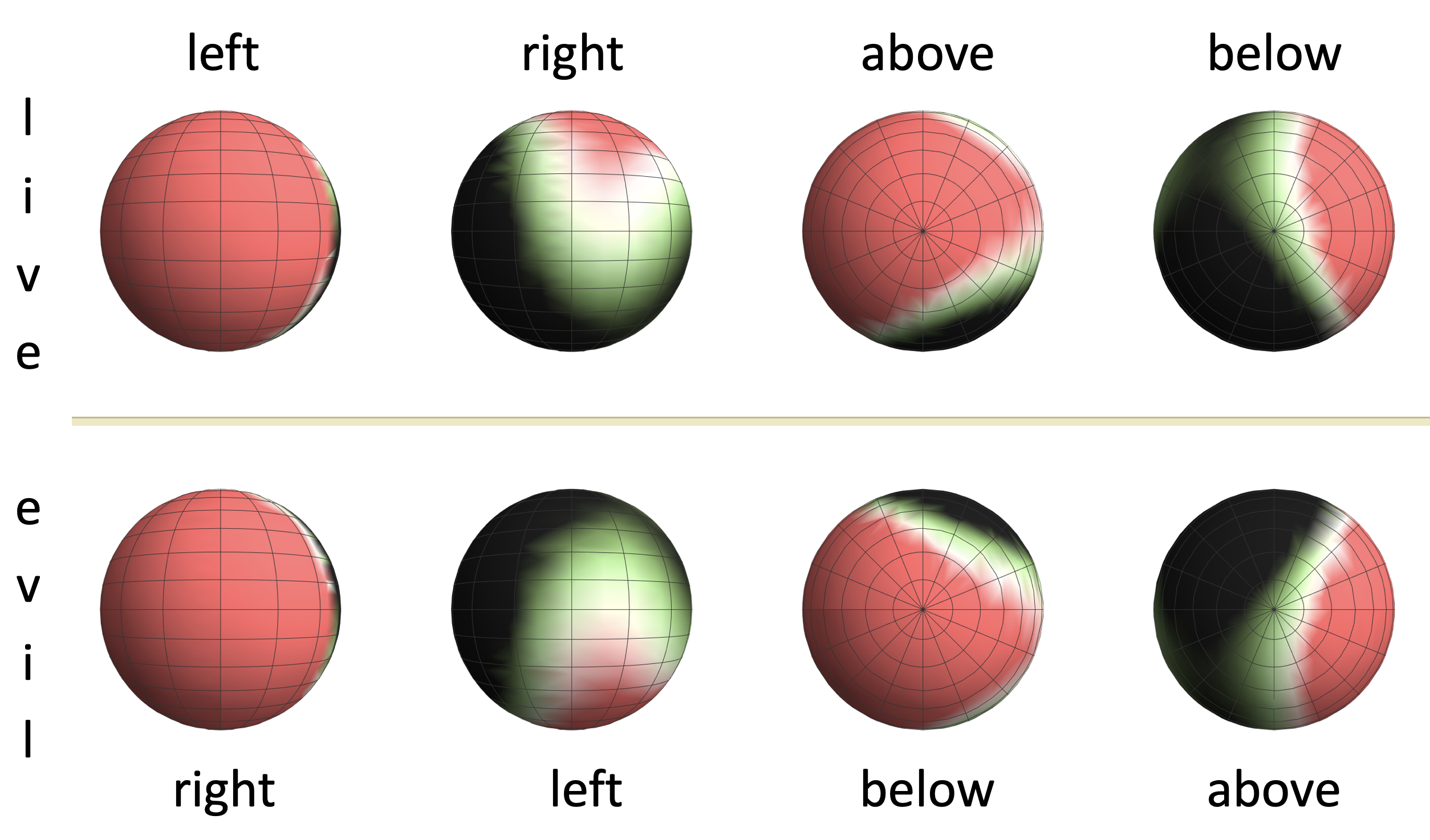}
\caption{Example of an electric charge distribution (projected on a sphere) of two biomolecules of opposite chirality, as seen from  left, from  right, from above and from below. This simply combines an electric dipole and an electric quadrupole. We need to reflect and rotate by 180 degrees the live molecule to obtain the evil one. The model we discuss in Section~\ref{sec-chiralpropa} is even simpler, combining electric and magnetic dipoles.}
\label{live_evil}
\end{figure}

\subsubsection{Surface Charge Density}
We describe the time-averaged surface charge density, $\Sigma$, on the sphere with a spherical harmonic expansion up to $l=2$, 
\begin{equation}\label{eq:surfcharge}
\Sigma=\frac1{4\pi R^2}\left(q+\frac{3{\bf d}\cdot{\bf r}}{R^2}+\frac{5\textbf r\cdot\textsf{Q}\cdot\textbf r}{2R^4}\right),\quad{\rm for}\,|\textbf r|=R,
\end{equation}
where \textbf r is a vector from the Carbon nucleus, $q=-4e$, $\textbf d$ is the electric dipole vector and $\textsf{Q}$ is the (symmetric, trace-free) quadrupole tensor. 

\subsubsection{Magnetic Dipole}
In addition, we suppose that there is a magnetic dipole \textbf m due to the motion of the electrons on the sphere, and neglecting the generally somewhat larger contribution associated with the spin of the binding electrons. The key assumption we make in this semi-classical model is that the magnetic dipole, as expressed by the magnetic field just outside the molecule is not perpendicular to \textbf d, as measured by the disposition of the electrons within the molecule. In other words, if for example, the electrons are more concentrated in the southern hemisphere, then they also flow preferentially in an equatorial direction. Assuming our model, the magnetic field within the molecule is uniform. 

\subsubsection{Molecular chirality}
As discussed in Sect.~\ref{sec:intro}, molecular chirality $\cal M$ has opposite signs for live and evil molecules. The charge density $\Sigma$ alone can exhibit electric chirality ${\cal M}_{\rm e}$. This is illustrated in Fig.~\ref{live_evil}, where $\Sigma$ is showed for two biomolecules of opposite chirality. However, we only consider in the following the simplest type of molecular chirality, the electromagnetic chirality ${\cal M}_{\rm em}$. (The electric quadrupole introduces additional effects which will be described elsewhere.)

\subsubsection{Electromagnetic Field}
The electric and magnetic fields associated with the charge density $\Sigma$ (Eq.~\ref{eq:surfcharge} with $\textsf Q=0$) and the magnetic dipole $\textbf m$ are: 
\begin{align}
\textbf E&=\frac1{4\pi\epsilon_0}\left(-\frac{q\textbf r}{r^3}-\frac{\textbf d}{R^3}\right),\;r<R;\quad=\frac1{4\pi\epsilon_0}\left(\frac{3(\textbf d\cdot\textbf r)\textbf r}{r^5}-\frac{\textbf d}{r^3}\right),\;r>R;\\
\textbf B&=\frac{\muup_0\textbf m}{2\pi R^3},\;r<R,\quad=\frac{\muup_0}{4\pi}\left(\frac{3(\textbf m\cdot\textbf r)\textbf r}{r^5}-\frac{\textbf m}{r^3}\right),\;r>R,
\end{align}

\begin{figure}
\centering
\includegraphics[scale=0.25]{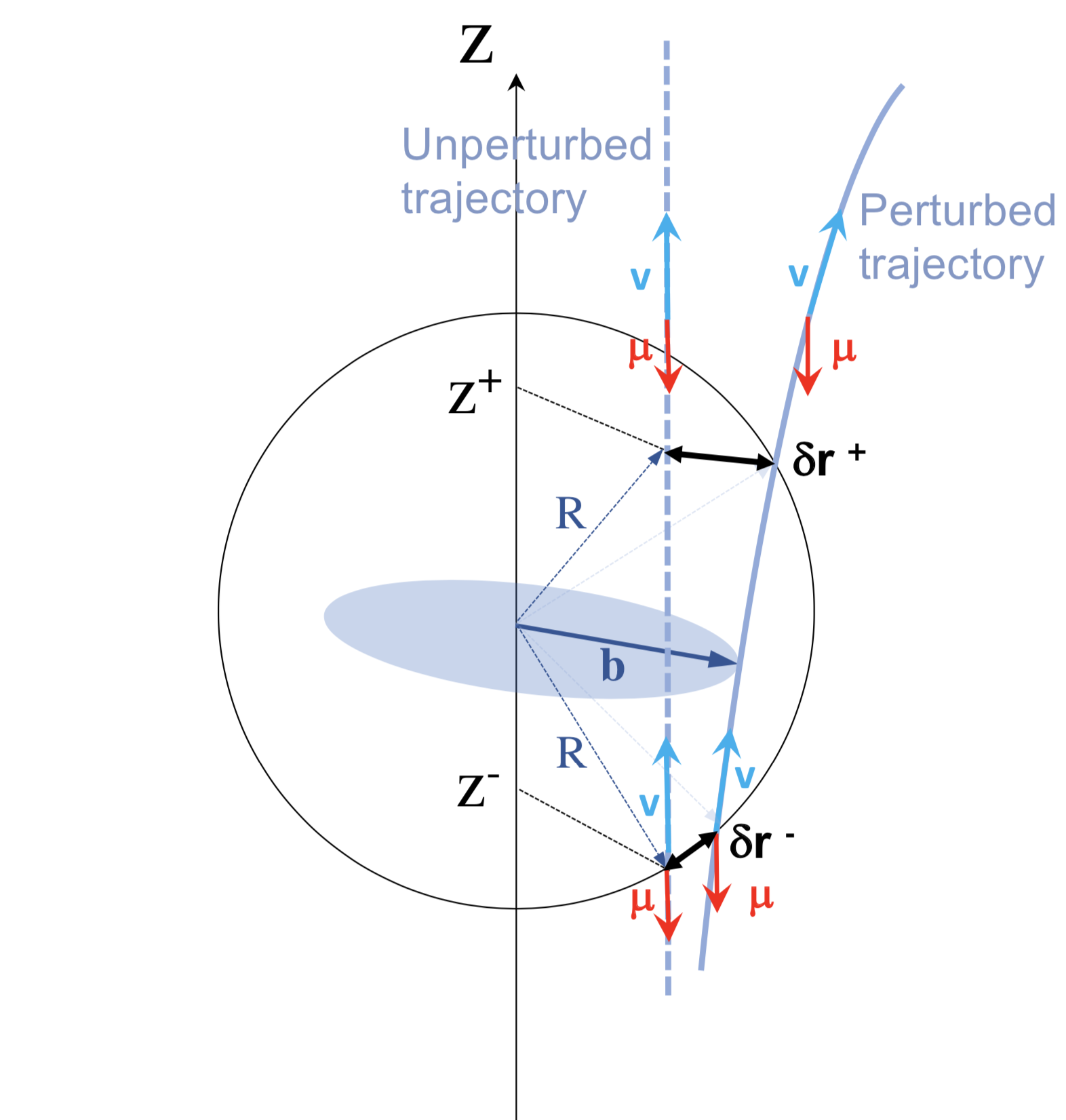}
\caption{Unpertubed vs. pertubed spin polarized cosmic-ray trajectories through a chiral unit. The unpertubed trajectory is along $z$. The perturbed trajectory due to the chiral molecular field $\textbf B$ is shown. The perturbed cosmic-ray therefore experience a slightly different charge distribution which would lead to a difference in the ionization rate between the two enantiomers.}
\label{sketch2}
\end{figure}

\subsection{Cosmic Ray Trajectory}
\subsubsection{Lorentz and Magnetic Dipole Forces}
We now consider the path of a single cosmic ray of charge $+e$, mass $M$, non-relativistic velocity \textbf v and impact parameter \textbf b with respect to the Carbon ion, as seen in Fig.\ref{sketch2}. For the moment, just consider one (live) molecule with a fixed, but arbitrary, orientation. The (classical) force acting on the cosmic ray is
\begin{equation}\label{eq:force}
\textbf F=e(\textbf E+\textbf v\times\textbf B)+\nabla\left[\boldsymbol{\muup}\cdot\left(\textbf B-\frac{\textbf v}{c^2}\times\textbf E\right)\right].
\end{equation}
The first term is the Lorentz force, which just depends upon the electric charge and combines electric monopolar, dipolar and magnetic dipolar components. 

The second term is the cosmic ray dipole force. It describes the net force exerted by the magnetic field, $\textbf B'=\textbf B-\textbf v\times\textbf E/c^2$, acting on the cosmic ray dipole moment $\boldsymbol{\muup}$ in the cosmic ray rest frame, to lowest order in $v$.

This interaction, which recalls elementary treatments of spin-orbit coupling and the Stern-Gerlach experiment, needs some discussion. If we treat the magnetic dipole, classically, as a small current loop and use the requirement that $\nabla\cdot\textbf B'=0$ in an inertial frame, then the form quoted is recovered. The force can be regarded as the spatial gradient of an interaction energy, $U_{\rm M}=-\boldsymbol{\muup}\cdot\textbf B'$. If we add the further requirement that the magnetic field in this frame is the gradient of a potential, which may not be true, then the alternative form, $(\boldsymbol{\muup}\cdot\nabla)\textbf B'$ is also obtained. This is what can be derived by treating the dipole as two equal, neighbouring, magnetic monopoles. However, the spin is quantum mechanical, not classical, and has to be included in a Hamiltonian formalism using $U_{\rm M}$. The form quoted in Eq.~\ref {eq:force} is therefore adopted.

\subsubsection{Magnetic Dipole Displacement}

We are interested in the interaction of the cosmic ray with the valence electrons as it enters and as it leaves the sphere. We introduce the coordinate $z=\textbf r\cdot\hat{\textbf v}$, with ${}\hat{}$ designating a unit vector, so that ingress and egress are at ${\textbf r}^\mp=\textbf b + z^\mp\hat{\textbf v}$ with $z^\mp=\mp(R^2-b^2)^{1/2}$. 
Let us start with the trajectory prior to ingress. We argued in Sec.~\ref{sec:intro} that the polarized, cosmic ray magnetic dipole can contribute a chiral preference. 
We also explained in Sec.~\ref{sssec:ESP} that the average spin polarization should be anti-parallel to the velocity.
We therefore  restrict attention to the force $\textbf F=\muup_{\rm z}\nabla B_{\rm z}$. Using $\nabla\times\textbf B=0$, we find that the velocity perturbation is $\delta\textbf v=\muup_{\rm z}\textbf B/Mv$. We are interested in the displacement perpendicular to the unperturbed trajectory at ingress not the displacement at a fixed time because the cosmic ray will eventually cross the sphere and it is the path the is important. This is given to first order by
\begin{align}
\delta\textbf r^-_\perp=&
\frac{\muup_{\rm z}}{2T}\int_{-\infty}^{z^-}dz\left(\textbf B-(\hat{\textbf v}\cdot\textbf B)\hat{\textbf v}\right),\nonumber\\
=&\frac{\muup_0\muup_{\rm z}}{8\pi TR^2}
\left[\left(\frac{1-(1+\eta)(1-\eta)^{1/2}}\eta\right)(\textbf m\cdot\hat{\textbf b})\hat{\textbf b}-\eta^{1/2}(\textbf m\cdot\hat{\textbf v})\hat{\textbf b}\right.\nonumber\\
&\left.-\left(\frac{1-(1-\eta)^{1/2}}\eta\right)(\textbf m\cdot\hat{\textbf v}\times\hat{\textbf b})\hat{\textbf v}\times\hat{\textbf b}\right].
\end{align}
where $T=Mv^2/2$ and $\eta=b^2/R^2$.

The velocity perturbation immediately after ingress has to take account of the impulse due to the current sheet in the sphere. However, there is no additional force as the interior magnetic field is uniform. The chiral part of the transverse displacement at egress can then  be shown to be 

\begin{align}
\delta\textbf r^+_\perp =&\delta\textbf r_\perp^-+\frac{(z_+-z_-)\delta\textbf v_\perp^+}v\nonumber\\
=&\frac{\muup_0\muup_{\rm z}}{8\pi T R^2}
\left[\left(\frac{1-(1+3\eta)(1-\eta)^{1/2}}\eta\right)(\textbf m\cdot\hat{\textbf b})\hat{\textbf b}-7\eta^{1/2}(\textbf m\cdot\hat{\textbf v})\hat{\textbf b}\right.\nonumber\\
&\left.-\left(\frac{1-(1+4\eta)(1-\eta)^{1/2}+6\eta(1-\eta)}\eta\right)(\textbf m\cdot\hat{\textbf v}\times\hat{\textbf b})\hat{\textbf v}\times\hat{\textbf b}\right].
\end{align}

\subsection{Ionization Rate}
\subsubsection{Cross Section}
The classical ionization cross section per Carbon is 
\begin{equation}
\sigma_{\rm ion}=16\pi a_0^2\left(\frac{I_H}I\right)\left(\frac{I_H}T\right)\sim3.5\times10^{-21}T_{\rm keV}^{-1}\,{\rm m}^2
\label{Ionization Cross_Section}
\end{equation} 
where $a_0\sim 0.5R$ is the Hydrogen Bohr radius. The probability that a cosmic ray incident upon an atom will create an ionization is therefore $P_{\rm ion}\sim\sigma_{\rm ion}/\pi R^2\sim0.2\,T_{\rm keV}^{-1}$. We take this ionization as a proxy for mutation. 

Direct measurements below $\sim1\,{\rm keV}$ give cross sections lower by factors up to ten and a slower decline with increasing kinetic energy \citep[{\textit e.g.}][]{Kim02}. This reflects the fact that more tightly bound electrons can be ionized as $T$ increases as well as quantum mechanical effects. Again, this is unimportant for our limited purpose.
In addition to its transverse displacement a cosmic ray will have a slightly different energy and cross section as it crosses the sphere and there is an associated chiral preference. This turns out to be subdominant in our model and we ignore it although it is likely to be significant in a more realistic description. 

\subsubsection{Second Order Perturbation to the Ionization Rate}
We have computed the first order deflection at ingress and egress. By itself, this leads to no net change in the ionization rate. However, the deflection results in the cosmic ray encountering a slightly different surface density of electrons and producing a change in the cosmic ray energy. The second order change in the relative ionization rate is then given by
\begin{eqnarray}
\delta\ln P_{\rm ion}&=&-\delta\textbf r_\perp^-\cdot\left(\nabla_\perp\ln\Sigma^- +\frac{e \textbf E^-_t}{T} \right) -\delta\textbf r_\perp^+\cdot\left(\nabla_\perp\ln\Sigma^+ +\frac{e \textbf E^+_t}{T}\right),\nonumber\\
&\approx&-(\delta\textbf r_\perp^-\cdot\nabla_\perp\ln\Sigma^- +\delta\textbf r_\perp^+\cdot\nabla_\perp\ln\Sigma^+),
\end{eqnarray}
where the perpendicular, logarithmic gradient in the relative surface charge density at ingress and at egress is
\begin{align}
\nabla_\perp&\ln\Sigma^\mp=\nonumber\\
&-\frac3{4eR^2}\left((1-\eta)^{1/2}\eta(\textbf d\cdot\hat{\textbf b})\hat{\textbf b}\pm\eta^{1/2}(\textbf d\cdot\hat{\textbf v})\hat{\textbf b}
+\textbf d\cdot\hat{\textbf v}\times\hat{\textbf b}\,\hat{\textbf v}\times\hat{\textbf b}\right),
\end{align}
after dropping the kinetic energy term as it is sub-dominant in our particular implementation.

\subsubsection{Averaging over Impact Parameter and Orientation}
So far, we have considered one molecule and one cosmic ray. The simplest assumption to make is that the cosmic ray flux is isotropic with respect to the molecule. This still allows the cosmic rays to be anisotropic if, as is likely, the molecules are randomly oriented, for example in water. (We emphasize that there are many circumstances when orientation biases may be present and these could lead to a larger chiral preference.)  

So, when a cosmic ray of fixed charge, mass, magnetic moment and speed encounters an ``atom'' described by $\textbf d,\textbf m$, we must average over the incident cosmic ray paths. In this case, any term in the $\delta\ln P_{\rm ion}$ sum that is odd in  $\hat{\textbf v}$ can be dropped as it will be canceled by the effect of a cosmic ray with the opposite velocity or impact parameter. We then average the remaining terms over azimuth, perpendicular to \textbf v using
$<\textbf m\cdot\hat{\textbf b}\;\textbf d\cdot\hat{\textbf b}>=\textbf m\cdot\textbf d/2$
etc and then average $\eta$ over a unit circle.

The final step is to average $\hat{\textbf v}$ over the surface of a unit sphere. Averages of the form $<\textbf u\cdot\hat{\textbf v}\,\textbf w\cdot\hat{\textbf v}>$ become $\textbf u\cdot\textbf w/3$. After performing these integrals and averages, we obtain
\begin{equation}\label{eq:probion}
\delta\ln P_{\rm ion}=\frac{3K\muup_0|\muup_{\rm e}||\textbf d||\textbf m|}{32\pi eTR^4}=1.3\times10^{-8}\left(\frac{d\,m}{1\,D\,\muup_{\rm B}}\right)\left(\frac T{1\,{\rm keV}}\right)^{-1},
\end{equation}
for a polarized positron encountering a chiral unit with \textbf d parallel to \textbf m. The constant $K$ evaluates to -3.06. $\muup_{\rm B}$ is the Bohr magneton.

\subsubsection{Chiral Preference}

We assume that the relative difference in the mutation rate $M$ between live and evil molecules is due to  the relative difference in the ionization rates, 
\begin{equation}
{\delta M}=\frac{M_{\rm evil}-M_{\rm live}}{M_{\rm live}}=\delta\ln P_{\rm ion}=f({\cal M},{\cal L},{\cal O}).
\label{demeter_eq}
\end{equation}
\begin{itemize}
   \item[{\footnotesize$\bullet$}] The cosmic-ray lodacity, $\cal L$, depends on the nature and energy of the primary cosmic rays, the external magnetic field $\textbf B_{\rm ext}$, and the nature of the medium where the cosmic ray shower develops. The lodacity allows for the depolarization of the cosmic rays as they lose energy;
   \item[{\footnotesize$\bullet$}] The molecular chirality, $\cal M$, depends on the nature of the chiral units and their relative configuration in the biomolecules (helix). For a single unit, it may includes all the possible different expressions of chirality, {\textit i.e.} ${\cal M}={\cal M_{\rm em}}+{\cal M_{\rm e}}$ (in our simple  model, ${\cal M}={\cal M_{\rm em}}=\hat{\textbf d}\cdot\hat{\textbf m}$);
   \item[{\footnotesize$\bullet$}] The molecular orientation, $\cal O$, takes account of orientation biases. It can depend on the presence of an external magnetic field, $\textbf B_{\rm ext}$, or a chiral material in the fount (like zeolites).
\end{itemize}

The effect that we have described is directly related to parity violation in the weak interaction. If we imagine a cosmic ray colliding with a molecule with impact parameter vector \textbf b and a mirror plane perpendicular to \textbf b passing through the central ion then an evil atom or molecule can always be rotated to be the reflection in this plane. If \textbf d is not perpendicular to \textbf m, a cosmic ray  with magnetic moment anti-parallel to its velocity will be deflected towards a region of greater charge density in one case and away from it in the other, leading to the difference in the mutation rate. (The non-zero value of $\textbf d\cdot\textbf m$ can be related to the volume integral of $\textbf E\cdot\textbf B=-Z_0^2\textbf d\cdot\textbf m/3\pi R^3$, where $Z_0$ is the impedance of free space.)  

The chiral preference calculated in this model is quite small. To order of magnitude, it is given by $\delta M\sim\alpha^2(m_e/M)(I/T)\sim10^{-10}$ for ten per cent polarized, mildly relativistic $e^\pm$ and $\sim10^{-11}$, for highly polarized mildly relativistic muons. Formally, $\delta M$ increases to $\sim10^{-5}$, for $e^\pm$ if $T\sim I$ but our semi-classical calculation is quite inappropriate and the loss of lodacity (Eq.~\ref{eq:LodT}) is likely to reduce $\delta M$ by a further factor of 300 to $3\times10^{-8}$. The presence of secondary electrons that are produced as the cosmic ray loses energy may also diminish the chiral preference.

A quantum mechanical calculation is necessary for $T\lesssim1\,{\rm keV}$ and must introduce additional spin- and charge-dependent effects as cosmic-ray electrons anti-symmetrize the wave function when there is an electron cosmic ray interacting with a valence electron and annihilation should be included for a cosmic ray positron. Models that better capture the helical nature of biological molecules and include the electric quadrupole, which can be very large \citep{Wu17}, are also likely to increase the chiral preference from that given by this estimate.  

\begin{figure}
\centering
\includegraphics[scale=0.35]{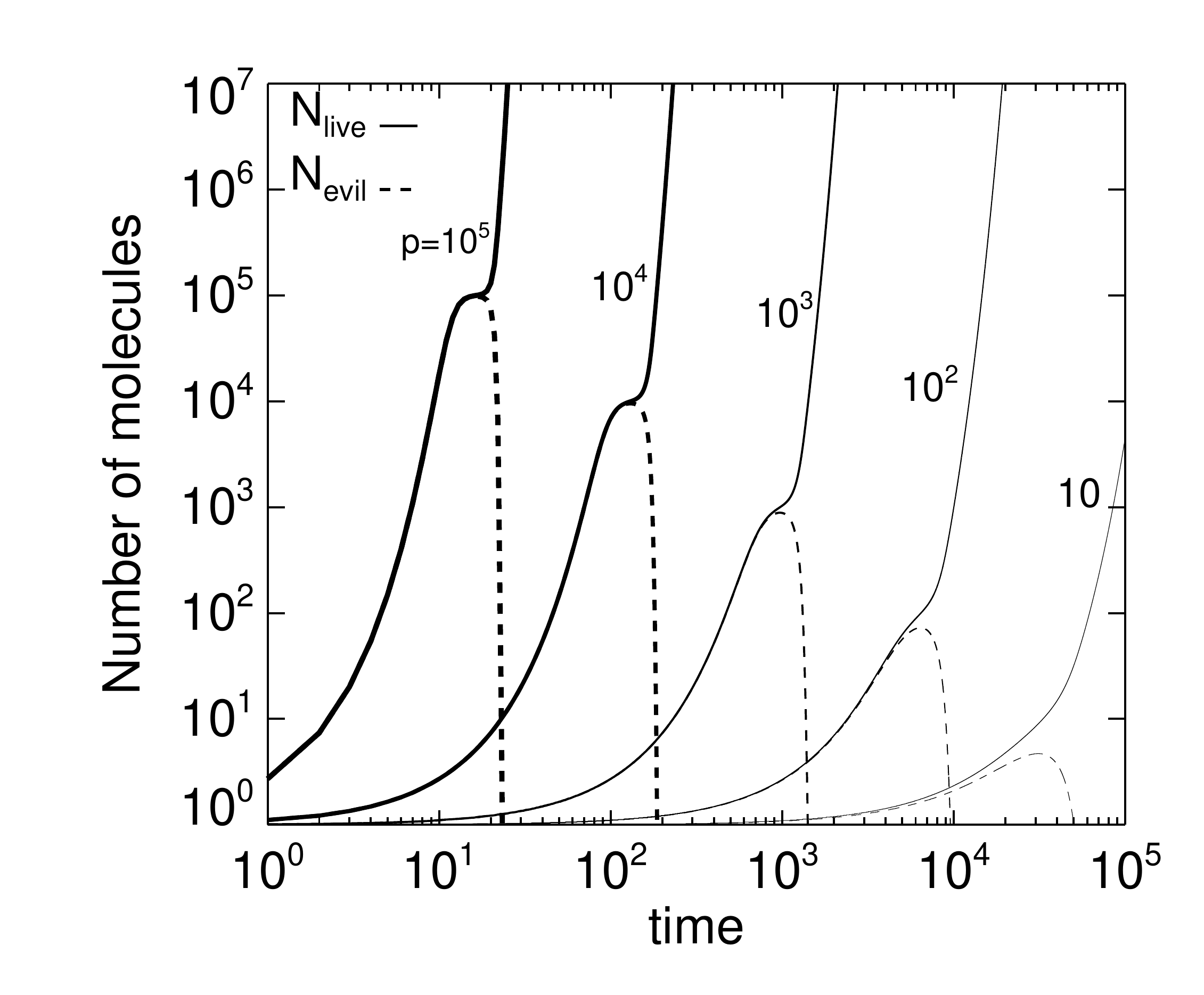}\\
\includegraphics[scale=0.35]{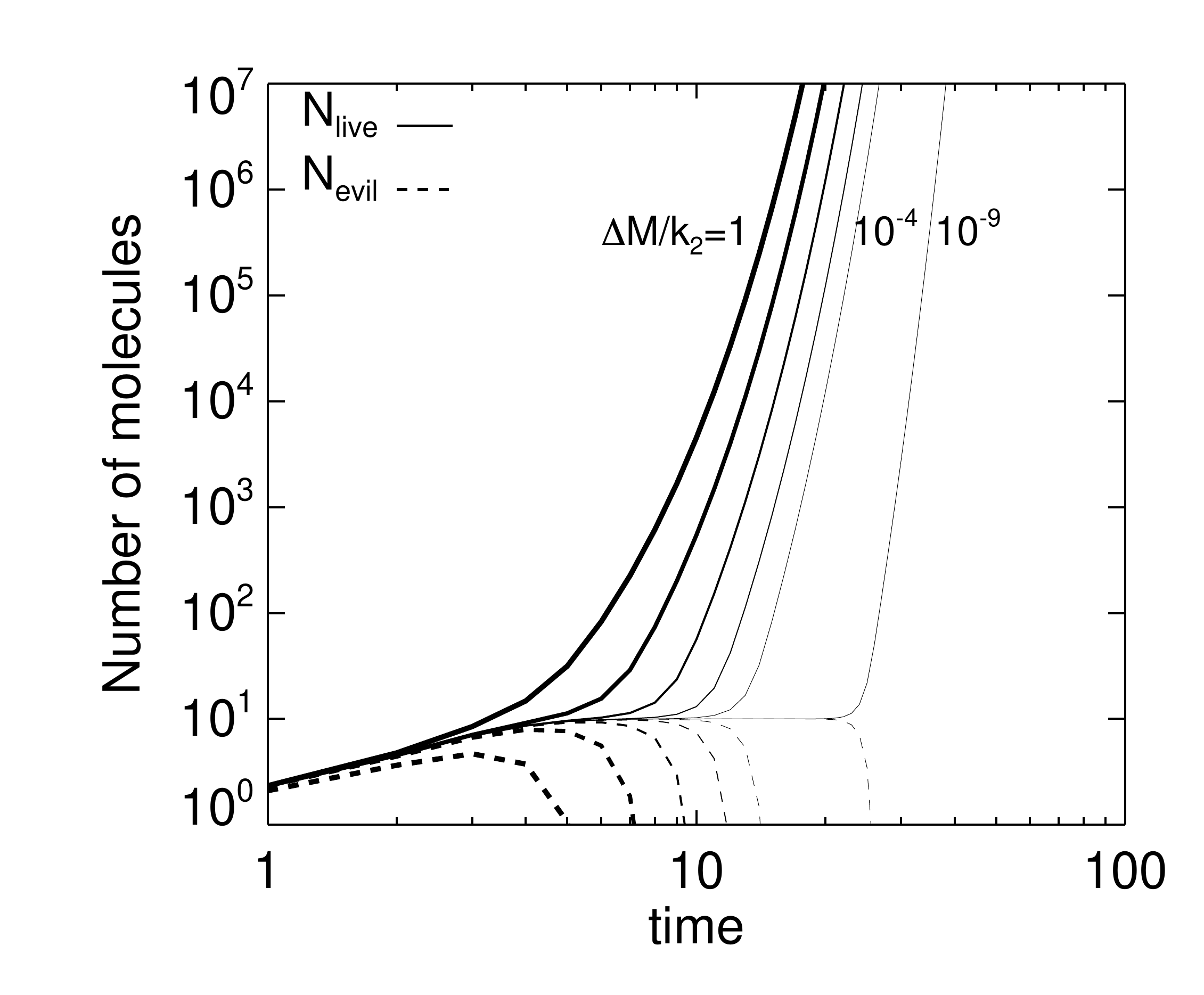}
\caption{Evolution of the number of live (solid lines) and evil (dashed lines) molecules, for different normalized proliferation rates  $p\equiv (k_1-M_{\rm live})/k_2=10^5$ to $10$ keeping $\Delta M=10^{-5}$ (upper panel);   different $\Delta M/k_2=1$ to $10^{-9}$ keeping $p=10$ (lower panel). The time is the normalized time $\tau=k_2 t$. We assume a racemic initial state $N_{\rm live}(t=0)=N_{\rm evil}(t=0)=1$. }
\label{evolution}
\end{figure}

\section{Evolutionary amplification}\label{sec-evolampli}

\subsection{From chiral preference to chiral dominance}
In this section, we show that the small differences in the propagation of cosmic rays through early biological molecules might be amplified by replication and reproduction, to lead to the chiral dominance we see today. The modeling of homochiralisation (the emergence of a single chiral life form) has been the object of various studies \citep[{\textit e.g.}][]{Frank53,Gleiser}. The existence of small chiral domains ({\textit i.e.} simple molecules of the same chirality) is a prerequisite for the assembly of structurally more stable polymers necessary for the formation of living systems; however, given the small chiral biases found in nature, it is difficult to explain the emergence of a single handedness with a pre-biotic process \citep{Bonner, BurtonBerger}. It has been also argued that the emergence of  homochirality  is a consequence of a small initial enantiomeric  excess amplified by the antagonism between the two mirror life forms \citep[][]{Frank53}. In this scenario, the proliferation rates of live and evil molecules are identical.
 However, we have shown in the previous section that the coupling between the cosmic-ray lodacity and the molecular chirality, expressed by the product  ${\cal L}\cdot {\cal M}$, leads to a difference in the mutation rate of live and evil molecules, which, as we show in the following, is sufficient to establish homochirality.

\subsection{Model equations}
In the following, we denote by $N_{\rm live}$ and $N_{\rm evil}$ the number of live and evil molecules, respectively.  
The evolution of the two populations  is the result of an interplay between stochastic factors (like spontaneous mutation) and deterministic factors (the cosmic radiation).
The proliferation rate of live and evil populations can be approximated as the difference between the natural growth rate, that we denote by $k_1$, and the mutation rate (natural and induced by the cosmic radiation) that we denoted by $M$. We denote by $k_2$ the antagonism factor.

A simple example of antagonism is the introduction of live monomers (naturally synthesized by  live forms) in the environment of the evil forms (and vice versa). It has been shown that endogenous right-handed amino acids  are used as auto-regulators, inhibiting bacterial growth under low nutrient conditions \citep{Cava11}. Therefore the introduction of wrong monomers (right-handed amino acids for live systems and left-handed amino acids for evil systems) will inhibit their respective growth. This is a simple example of antagonism; other biological effects could be important but our purpose is only to show a qualitative effect.

We assume that we can neglect the natural mutation (as it is the same for live and evil molecules) and therefore, we define $M_{\rm live}=\sigma_{M,{\rm live}} \Phi$ and $M_{\rm evil}=\sigma_{M,{\rm evil}} \Phi$, where $\sigma_M$ is the mutation induction cross section (in $\mu m^2$) given by Eq.~\ref{mutation_cross_section} and $\Phi$ is the particle flux (in $\mu m^{-2}s^{-1}$). 
The exposure to the cosmic-rays radiation is a deterministic factor because it induces a consistent chiral bias in the form of a difference in the  mutation rate $\delta_M=(M_{\rm evil}-M_{\rm live})/M_{\rm live}$ (see Eq.~\ref{demeter_eq}). As it has been shown, $\delta_M$ is directly related to the sign and amplitude of the molecular chirality $\cal M$, the lodacity ${\cal L}$, and other factors like the molecular orientation.   The factor $k_1$ and $k_2$ are  independent of the chirality of the molecules.

The evolution of the two populations is described by the logistic equations:
\begin{eqnarray}
    \frac{d\ln N_{\rm live}}{dt}&=k_1-M_{\rm live}-k_2N_{\rm evil}\,,\label{logistic}\\
    \frac{d\ln N_{\rm evil}}{dt}&=k_1-M_{\rm evil}-k_2N_{\rm live}\,.
    \label{logistic2}
\end{eqnarray}

Apart from $\delta M$ that we have estimated  in our simple model, the other factors are unknown; we can approximate $k_1$ as the growth rate of bacteria on Earth. But we are left with another unknown factor, the conflict rate between live and evil molecules. Therefore we normalize the equations by the conflict rate $k_2$:

\begin{eqnarray}
    \frac{d\ln N_{\rm live}}{d\tau}&=&\frac{k_1-M_{\rm live}}{k_2}-N_{\rm evil}\,,\label{logistic}\\
    \frac{d\ln N_{\rm evil}}{d\tau}&=&\frac{k_1-M_{\rm live}}{k_2}-\frac{\Delta M}{k_2}-N_{\rm live}\,,
    \label{logistic3}
\end{eqnarray}
We introduce the normalized proliferation rate $p\equiv (k_1-M_{\rm live})/k_2$, the normalized time $\tau=k_2 t$  and $\Delta M \equiv M_{\rm evil} - M_{\rm live}$. Note the relation $\Delta M= \delta_M {M_{\rm live}}$.

\subsection{Illustrative solutions}
 Here we show the effect of varying the strength of the parameters $k_1$, $k_2$ and $\Delta M$  in the evolution of the two populations.
Solutions to the Eqs.~\ref{logistic}-\ref{logistic2} are presented in Fig.~\ref{evolution}. We assume that we start at $t=0$ with a distribution of live and evil molecules that is racemic in average (for simplicity we assume $N_{\rm live}=N_{\rm evil}=1$).
The time needed to reach a pure homochiral state depends on the normalized parameters $p$, $\Delta M/k_2$.  To illustrate the effect of these different parameters, we show in Fig.~\ref{evolution} the evolution of the population number as a function of time, varying $p$ (upper panel) and $\Delta M/k_2$ (lower panel). As we can see qualitatively from these curves, the cosmic-rays assure a consistent chiral bias allowing a difference in the evolutionary path of live and evil systems, while the selection pressure term determines the time scale at which homochirality can be established. 

The key question, that we cannot answer because we cannot make reliable estimates, is whether the natural growth rate, $k_1$ and the antagonism factor $k_2$ are sufficient to transform a small chiral preference into chiral dominance. However, we note that simple bacteria can replicate in hours and the time scale for environmental evolution on a young planet is likely to be millions of years. There could be many billions of generations to allow the processes described to take effect. 

\section{Discussion}\label{sec-discussion}
In this paper, we have argued that homochirality is a deterministic consequence of the weak interaction, expressed by cosmic irradiation of molecules as they transition from pre-biotic to vital components of single-celled organisms. The choice that was made is then traceable to the preponderance of baryons over antibaryons, established in the early universe and ultimately to the symmetries of fundamental particle interactions presenting requirements as first elucidated by  \citet{Sakharov67}. In a similar fashion, we have tried to list some of the physical and chemical factors that seem to be necessary for such a causal biological path to have been followed. In addition we have introduced one specific mechanism, involving collisional ionization by secondary spin-polarized cosmic rays, that, we argue, is likely to be more relevant than mechanisms involving circularly polarized ultraviolet light. We have also demonstrated how a quite small chiral preference can evolve into chiral dominance, emphasizing the importance of conflict. Much more study is needed to determine if these ingredients do, indeed, suffice to account for homochirality or if, instead, the choice must be due to chance or environmental idiosyncrasy. Answering this question is central to understanding the origin, prevalence and migration of life in the universe.

It should be emphasized that the purpose of this calculation is not to obtain a quantitative answer but, instead, to demonstrate the qualitative effects involved and to highlight the various factors that have to be included. In particular, our restriction to the $\textbf d\cdot\textbf m$ interaction surely underestimates the molecular chirality, because actual bonds have contributions from higher $\ell$ multipoles which lead to larger gradients in both the magnetic field and the charge density. For example, it has been shown that the adsorption of chiral molecules on specific surfaces can substantially enhance the optical activity by several orders of magnitude ($\sim1000$) because of the electric dipole - electric quadrupole interaction \citep[{\textit e.g.}][]{Wu17}, {\textit i.e.}, the electric chirality ${\cal M_{\rm e}}$ that we neglected so far.  
We envision that, under specific conditions, ${\cal M_{\rm e}}$ would lead to an enhancement of the enantioselectivity in our model.

If we admit a deterministic path for life, then life's handedness is a worldwide property (here "wordwide" has to be taken in Giordano Bruno' sense in which our world is only one of many worlds). So far,   enantiomeric excesses of left-handed amino acids were found in  a very limited sample of carbonaceous chondrites  subject to terrestrial contamination. The search for enantiomeric excesses in amino acids {\it in situ}  is the goal of future  space missions. The Hayabusa2 \citep{Hayabusa2} and OSIRIS-REx \citep{OSIRIS} spacecrafts will return to Earth with samples in 2020 and 2023, taken from two small carbon-rich asteroids. The future mission ExoMars 2020 \citep{ExoMars} 
is planned  to return to the Earth with samples collected from Mars. The main goal of these missions is to determine the sign of enantiomeric excesses in the chiral molecules found in these samples.
Based on the results from pre-biotic mechanisms, we do not expect to find large  enantiomeric excesses of amino acids in places where living systems were absent. However, according to our model,  the sign of the small excesses is expected to be the same everywhere where cosmic-rays showers can develop.

A way to test the proposed scenario and further our understanding of the processes we have highlighted is to perform experiments. The mutation rate can be  estimated through several methods  \citep[\textit{e.g.}][]{Foster06,Pope08}. A prediction of our model is that the mutation rate is dependent upon the spin-polarization of the  radiation. One possible experiment would be to measure the mutation rate of two cultures of bacteria under spin-polarized radiation (either  $e^\pm$ or $\mu^\pm$) of different lodacity with energy above the threshold necessary to induce double strand breaks in DNA ($\sim50$ eV). If the coupling between lodacity and molecular chirality is efficient in introducing a chiral bias, one of the two cultures should exhibit a much lower mutation rate. We emphasize that much  can be learned experimentally from the comparison of chiral molecules involved in biology and using both signs of lodacity which can be created at accelerators. It is not necessary to create ``mirror life'' to proceed. Once the dominant processes are identified, we can have confidence in our understanding of particle physics and quantum chemistry to draw the necessary conclusions.

If these experiments show that the evolution of  bacteria is influenced by the spin-polarized radiation, this will be a good indication that spin-polarized cosmic-rays might be an important piece of the chiral puzzle of life. 

\section*{Acknowledgements}
The research of NG is supported by the Koret Foundation, New York University and the Simons Foundation. NG thanks Louis d'Hendecourt for helpful discussions.

\onecolumn
\appendix

\end{document}